\documentclass[aps,prb,twocolumn,superscriptaddress,showpacs]{revtex4-1}
\usepackage{amsmath,graphicx,color}
\begin{document}
\title{Transversal spin freezing and re-entrant spin glass phase in chemically disordered Fe-contained double perovskite multiferroics }
\author{V.A.Stephanovich}\email{stef@uni.opole.pl}
\affiliation{Institute of Physics, Opole University, Opole, 45-052, Poland}
\author{V.V.Laguta}\email{laguta@fzu.cz}
\affiliation{Institute of Physics, Opole University, Opole, 45-052, Poland}
\affiliation{Institute of Physics AS CR, Cukrovarnicka 10, Prague, 16200, Czech Republic}

\begin{abstract}
We propose experimental verification and theoretical explanation of magnetic anomalies in the complex Fe-contained double perovskite multiferroics like PbFe$_{1/2}$Nb$_{1/2}$O$_3$. The theoretical part is based on our model of coexistence of long-range magnetic order and spin glass in the above substances. In our model, the exchange interaction is anisotropic, coupling antiferromagnetically $z$ spin components of Fe$^{3+}$ ions. At the same time, the $xy$ components are coupled by much weaker exchange interaction of ferromagnetic sign. In the system with spatial disorder (half of corresponding lattice cites are occupied by spinless Nb$^{5+}$ ions) such frustrating interaction results in the fact that antiferromagnetic order is formed by $z$ projection of the spins, while their $xy$ components contribute to spin glass behaviour. Our theoretical findings are supported by the  experimental evidence of coexistence of antiferromagnetic and spin glass phases in chemically disordered Fe-contained double perovskite multiferroics.
\end{abstract}

\date{\today}
%\pacs{73.20.-r, 73.40.-c, 73.21.Ac, 72.80.Tm}

\maketitle

\section{Introduction}

In the context of novel materials with unusual physical properties, the researchers are interested in the Fe-based double perovskite multiferroics with the general formula PbFe$_{1/2}$M$_{1/2}$O$_3$ (M=Nb, Ta, Sb)  and their solid solutions with substitution of A or B type ions in the ABO$_3$ perovskite structure, see, e.g. \cite{klem2010,kumar08,rotaru09} and references therein.
Recent studies \cite{singh10,singh08} of these substances reveal a lot of interesting properties like large magnetoelectric coupling and high dielectric permittivity. It had been shown in the above papers that these properties occur in substantial range of temperatures and chemical compositions, revealing the existence of ferroelectric (FE), antiferromagnetic (AFM) and spin glass (SG) phases on the corresponding phase diagrams. In the above compounds, Fe$^{3+}$ and M$^{5+}$ cation positions may be ordered or disordered within the simple cubic B sublattice of the perovskite ABO$_3$ structure. The degree of chemical ordering depends on the relative strengths of electrostatic and elastic energies and on the ionic radii of these cations in particular. It is commonly accepted that Pb(Fe$_{1/2}$Nb$_{1/2}$)O$_3$ (PFN) and Pb(Fe$_{1/2}$Ta$_{1/2}$)O$_3$ (PFT) are chemically disordered compounds due to almost equal ionic radii of Fe$^{3+}$ and Nb$^{5+}$ or Ta$^{5+}$ \cite{shannon76}, while Sb-contained compounds can be chemically ordered up to 90\% as Sb$^{5+}$ is much larger than Fe$^{3+}$ \cite{sib13,my14}. The magnetism of the compounds is due to Fe$^{3+}$, S = 5/2 ions that occupy half of octahedral sites of the perovskite lattice. The magnetic moments of the Fe$^{3+}$ ions interact with each other via various superexchange paths, considered in Ref. \onlinecite{kuzian14} in details. 

The majority of papers consider the spin glass state as the magnetic ground state of both PFN and PFT at $T<T_g\approx10-15$ K. There are several ambiguous statements about SG nature of the magnetic ground state in PFN at $T < 20$ K, see \cite{klem2010} and references therein. The statement about glasslike state, starting at $T=120$ K for low magnetic fields $H=100$ Oe or at $T=28$ K at $H \geq 1000$ Oe \cite{kumar08} along with reference to some superparamagnetic (SPM) behavior with blocking temperature $T_B$ increase the confusion in understanding of the above magnetic ground state nature. The light was poured in the paper \cite{rotaru09} with the help of $\mu$SR spectroscopy and neutron scattering. The authors \cite{rotaru09} have shown that magnetic ground state of PFN is a spin glass like state, that coexists with the long-range AFM order below $T_g \approx 20$ K in the time scale of their experiment. The SG state has also been identified from $^{17}$O NMR as distinct anomalies in the spin-lattice and spin-spin nuclear magnetic relaxation \cite{blinc08}. However, the microscopic nature of the above SG state as well as essential increase of magnetic susceptibility in PFN and PFT below the Neel temperature remain unclear till now. It has been proposed in Refs. \onlinecite{klem2010} and  \onlinecite{laguta13} that along with infinite-range percolation cluster responsible for the long-range ordered AFM phase, superantiferromagnetic Fe$^{3+}$ clusters are created also. The latter are responsible for the spin-glass like (so-called cluster glass) behavior of magnetic properties. In principle, this fact agrees with NMR and ESR results \cite{laguta13, laguta10}. $^{93}$Nb NMR spectra in PFN \cite{laguta10} show the existence of two different Nb sites with different local magnetic fields: Fe-rich, Nb-poor and Fe-poor, Nb-rich nanoregions. These data suggest that a spin-glass state of PFN below 11 K might arise from the latter regions and a phase separation exists, at least, at nanometric scale. The second model, recently proposed in Ref. \onlinecite{chilal13}, is based on coexistence of the long-range order and SG on the microscopic scale. It assumes that all Fe$^{3+}$ spins in the system form AFM order below the Neel temperature, but there are additional long-range spin-spin correlations along $z$ direction, while the transversal $xy$ spin components undergo random thermal reorientations between energetically equivalent (or nearly equivalent) orientations. It has been suggested that such system of Heisenberg spins gradually froze into a SG state, known as $T_{xy}$ reentrant SG phase \cite{ryan92}.  However, the theoretical description of such reentrant phase is absent so far for PFN or PFT so that the microscopic origin of this state still remains unknown. 

The detailed magnetoelectric studies of PFN single crystals have been performed in Refs. \onlinecite{klem2010,watan89}. In particular, it had been found \cite{watan89} that below $T_c^m \approx 9$ K the lattice point group symmetry changes from $m_x1'$  to $m'_x$.  It was concluded therefore that a weak ferromagnetism, usually observed in PFN, is induced in an originally antiferromagnetic spin structure by lowering the crystal symmetry. This finding increase the confusion in understanding of magnetic ground state nature of both PFN and PFT.

The aim of this paper is to make regimentation of the diverse (and sometimes controversial) facts about the coexistence of long-range magnetic order and spin glass phase in the above double perovskite multiferroics. For that, based on so-called random local field method (see \cite{ss02,ss03} and references therein) we are going to present the theoretical description of the mixed AFM-SG phase in the perovskite multiferroics. Besides we present strong experimental evidence of such coexistence.

\section{Theoretical approach}
\subsection{Qualitative considerations}

\begin{figure}
\begin{center}
\includegraphics [width=0.47\textwidth]{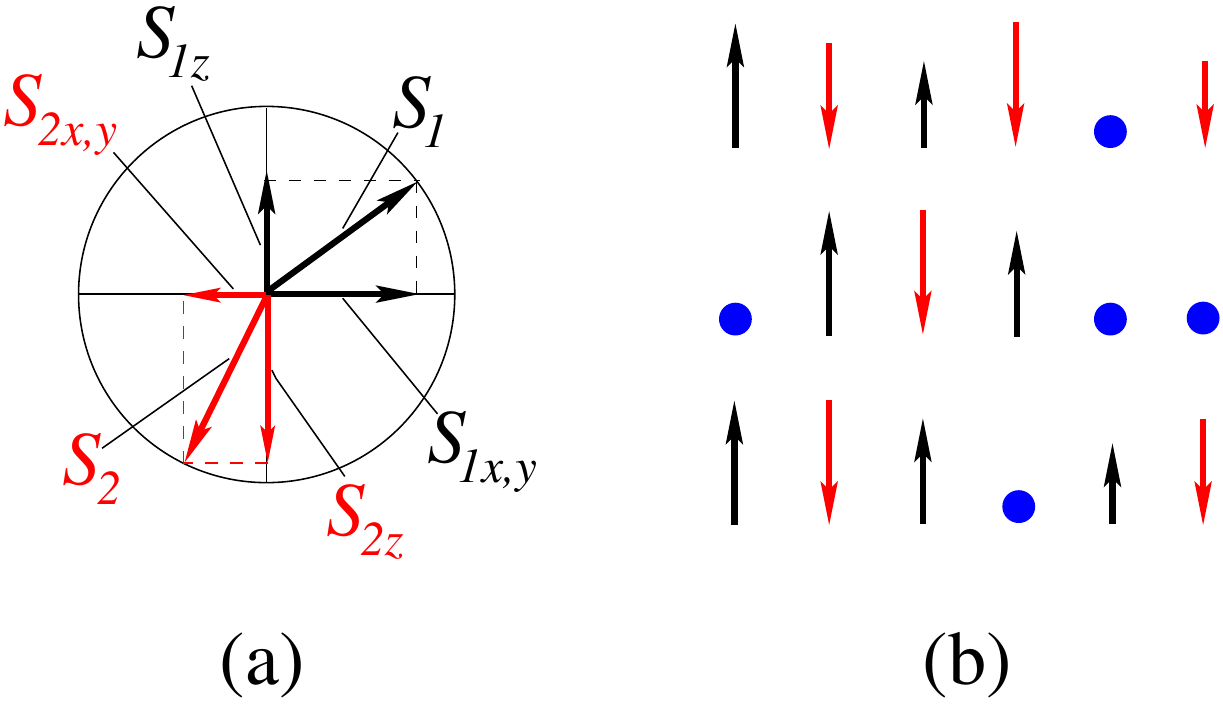}
\end{center}
\caption{(color online). The explanation of spin $z$ component formation (a) and sketch of Fe$^{3+}$ spins $z$ components AFM alignment in PbFe$_{1/2}$Nb$_{1/2}$O$_3$ (b). Black and red arrows show the magnetic sublattices in AFM alignment and blue circles on panel (b) show the spinless Nb$^{5+}$ sites. The lengths of spin vectors ${\bf S}_1$ and ${\bf S}_2$ are the conserved quantites, their $x$ and $y$ components $S_{1,2x,y}$ fluctuate from site to site so that the lengths $S_{1,2z}$ are different.}\label{Fig1a}
\end{figure}

The main peculiarities of above perovskites, making them different from ordinary antiferromagnets are the sharp increase of magnetic susceptibility in the antiferromagnetic phase $T<T_N$ with its subsequent diminishing at low temperatures $T<T_g$, where $T_N$ and $T_g$ are, respectively, Neel and glassy transition temperature. In this section we are going to show that these anomalies can be well described within our model of mixed AFM-SG phase, which is realized in the PFN and PFT.  It has been demonstrated experimentally in Ref. \cite{chilal13} that SG and AFM phases coexist in PFN on the microscopic scale. The crux of the matter is that in ground and low-lying excited states of any magnet the length of its magnetization vector (or the lenghts of sublattice magnetizations in AFM) is conserved, 
i.e. that vector can only rotate, keeping its length constant, see Fig. \ref{Fig1a} (a). At the same time if we assume that the interplay between disorder (random positions of magnetic Fe$^{3+}$ ions and nonmagnetic Nb$^{5+}$ ions) and anisotropic spin-spin interaction makes $x,y$ spin components fluctuate so that (by virtue of conservation of spin vector length) $z$ spin components, which are antiferromagnetically aligned, have different lengths. This means that while Fe$^{3+}$ spins in PFN form AFM order along $z$ -axis (Fig. \ref{Fig1a} (b)), their $xy$ components contribute to SG phase. 

For this scenario to realize, the interaction between spins should have several contributions. Namely, although there are short-range exchange and superexchange interactions, generating long-range (AFM in our case of PFN) magnetic order, there is one more type of interaction, inevitably present in any magnetic system and fixing the direction of its magnetization. This is so-called relativistic spin-spin interaction, forming magnetic anisotropy energy and magnetic dipole interaction. Although the amplitudes of latter interactions are usually (much) smaller then that of exchange interaction, they play an important role, being responsible both for spontaneous magnetization (and/or antiferromagnetic vector) direction as well as for features like magnetic domain wall width. As the magnetic dipole interaction depends on the angle between spins, its action in the disordered spins system leads to their gradual freezing in random orientations, yielding spin glass state. Moreover, the freezing temperature, $T_g \approx 11 $ K is much smaller then Neel temperature, $T_N \approx 150$ K as the resulting exchange interaction is accordingly weaker then initially AFM one. In other words, $T_N$ is determined by the AFM exchange interaction amplitude, while $T_g$ - by the synergy of AFM and FM interactions. 

Here we argue that the main reason for above behaviour is the synergy between disorder and the presence of several types of interactions between Fe$^{3+}$ spins in PFN. Namely, below we assume that there is short-range (of range $z_0$) exchange interaction of AFM sign along $z$ axis, while in the $xy$ plane there is so-called frustrating exchange interaction of FM sign (as opposed to AFM one in 
$z$ - direction), which contributes to SG behavior, rising $T_g$ and lowering $T_N$. 

Note that the consideration of the sole chemical Fe - Nb disorder in PFN does not explain the values of $T_N$ and $T_g$ temperatures in it. Really, consider perovskite ABO$_3$ ferrites without above disorder, i.e. the systems like YFeO$_3$, where all B lattice sites are occupied by Fe. If we imagine the dilution of such systems by nonmagnetic ions like in PFN and extrapolate (linearly with Fe concentration) their $T_N$, which is around 600 K (see, e.g., Ref. \onlinecite{yfo}) to, say, 50\% dilution, we obtain $T_N \sim $ 300 K, i.e. much larger value then $T_N \sim $ 150 K for PFN. Our theory explains this and other experimental facts by considering the reciprocity between the chemical disorder and frustrating FM - AFM exchange interactions, leading to glassy type of spin-spin correlations at low temperatures.  

In principle there is also long-range magnetic dipole interaction of relativistic nature, but its amplitude is negligibly small in PFN \cite{kuzian14}. According to estimations of Ref. \onlinecite{kuzian14}, the AFM exchange interaction between nearest Fe$^{3+}$ is around 42 K, while the next-nearest neighbour FM interaction constant is $\sim 1K$. At the same time, the constant of magnetic dipole interaction is around 0.077 K, which is more then one order of magnitude less then above smaller exchange constant.  Below we use these values of exchange constants to fit the experiment in PFN.

As the length of spin vector is usually conserved, in our picture (see Fig.\ref{Fig1a}) the fluctuations of the spin is due to those of its $xy$ projections, which means that $z$ - projection is always directed along $z$ axis but its length fluctuates due to those of $xy$ projections. Note that in this case the average balance between $+z$ and $-z$ directions of $S_z$ is conserved so that the overall structure is antiferromagnetic. 

To describe the above discussed effects quantitatively, we introduce so-called random field model, which includes naturally the explicit form of the interaction between the Fe$^{3+}$ spins in PFN and PFT.

\subsection{Random local field model} 
The Hamiltonian of the ensemble of spins 5/2 reads
\begin{equation}\label{rf1}
{\cal H}=-\sum_{ij}J_{ij}{\bf S}_i{\bf S}_j+g\mu_B H_{ext}\sum_i S_{iz}, 
\end{equation}
where ${\bf S}_i$ is Heisenberg spin in $i$ - th host lattice site, $H_{ext}$ is an external magnetic field, directed along $z$ axis, $J_{ij}=$ $J({\bf r}_{ij})$ (${\bf r}_{ij}={\bf r}_{j}-{\bf r}_{i}$) is a spin-spin interaction, which, in the spirit of above discussion, we choose in the form
\begin{eqnarray}
&&J({\bf r})=-J_{zz}\exp(-z/z_0)+\nonumber \\
&&+J_{xy}\exp(-\sqrt{x^2+y^2}/R_0). \label{rf2}
\end{eqnarray}
Here summation is running over host lattice sites, where Fe$^{3+}$ ions (i.e. spins) are present. We consider the strengths of the interactions $J_{zz}$ and $J_{xy}$ to be close to those from Ref. \onlinecite{kuzian14}, but they may be adjusted to achieve the better fit to experiment. For simplicity, the Hamiltonian \eqref{rf1} does not contain so-called zero field splitting terms, related to single ion anisotropy.

Hamiltonian \eqref{rf1} incorporates two sources of randomness. The first is the spatial disorder, which means that spin can be randomly present or absent in the specific $j$-th cite of a host lattice. The second is the thermal disorder, i.e. random spin projection in the $j$-th cite. Having this randomness in mind, we consider every spin ${\bf S}_j$ as a source of a random field ${\bf H}_{ri}=\sum_{j\neq i}J_{ij}{\bf S}_j$ affecting other spins at the sites ${\bf r}_i$. In other words, every spin in our approach is subjected to some random, fluctuating field, created by the rest of the spin ensemble. Thus all thermodynamic properties of the system are determined by the distribution function $f({\bf H}_r)$ of the random field ${\bf H}_r$. That is, any spin-dependent macroscopic quantity $<<B>>$ (like magnetization), reads $<<B>>=\int <B>_{\bf H}f({\bf H})d^3H$, where ${\bf H}\equiv {\bf H}_r$ and
$<B>_{\bf H}$ is auxiliary single particle thermal average with effective Hamiltonian ${\cal H}_{eff}=\sum_i({\bf H}_i+{\bf H}_{ext}){\bf S}_i$.

The explicit form of distribution function $f({\bf H})$ reads
\begin{equation}\label{rf3}
f({\bf H})=\left\langle \overline{\delta\left({\bf H}-\sum_{j\neq i}J_{ij}{\bf S}_j\right)}\right\rangle,
\end{equation}
where bar denotes the averaging over random spatial positions of spins (spatial averaging) and angular brackets denote the thermal averaging over possible spin orientations. To actually perform the above averagings, we use the spectral representation of $\delta$ - function. \cite{ss02,ss03,st97}
As we cannot do these averagings exactly \cite{ss02,ss03,st97}, we do them self-consistently in the framework of statistical theory of magnetic resonance lineshape. \cite{stoneh} For the general form of interaction $J({\bf r})$, incorporating not only $J_{zz}$ and $J_{xy}$, Eq. \eqref{rf2}, this procedure yields
\begin{widetext}
\begin{subequations}\label{r4}
\begin{eqnarray}
&&f({\bf H},{\bf M})=\frac 1{(2\pi)^3 }\int d^3\rho\ \exp \left\{ i(\rho_x H_x+\rho_y H_y+\rho_z H_z)+ n\int _V\left[\frac{\sinh 3\gamma}{\sinh(\gamma/2)}\frac{\sinh (\phi^{-1}(M)/2)}{\sinh 3\phi^{-1}(M)}-1\right]d^3r\right\}, \label{rf4a} \\
&&\gamma=\Biggl\{\bigl[\phi^{-1}(M)\bigr]^2-(Q_x^2+Q_y^2+Q_z^2)+
2i\biggl[Q_xM_x+Q_yM_y+Q_zM_z\biggr]\Biggr\}^{1/2},\ M=\sqrt{M_x^2+M_y^2+M_z^2}, \nonumber \\
&&Q_x=J_{xx}\rho_x+J_{xy}\rho_y+J_{xz}\rho_z,\ Q_y=J_{xy}\rho_x+J_{yy}\rho_y+J_{yz}\rho_z, \
Q_z=J_{xz}\rho_x+J_{yz}\rho_y+J_{zz}\rho_z, \nonumber \\
&&M_{x,y,z}=\int \frac{H_{x,y,z}}{H}\ B_{5/2}\left(\frac 52 g\beta \mu_B H\right)f({\bf H},{\bf M})d^3H. \label{rf4b}
\end{eqnarray}
\end{subequations}
\end{widetext}
The equations \eqref{rf4a} and \eqref{rf4b} constitute a self-consistent set for determination of the dimensionless magnetization components $M_{x,y,z}={\cal M}/M_0$ ($M_0$ is saturation magnetization) within random local field model. We pay attention that as we are dealing with antiferromagnet, consisting of two sublattices with opposite directions of magnetizations, there are two types of magnetization in AFM. One is above magnetization ${\bf M}=\sum_i {\bf S}_i$ and the other is staggered magnetization (or so-called AFM vector) ${\bf L}=\sum_i e^{i{\bf Q}{\bf r}_i}{\bf S}_i$, where ${\bf Q}=(\pi,\pi,\pi)$ is ordering wavevector (see, e.g. Ref. \onlinecite{qm} and references therein). The equations for ${\bf L}$ then will be almost similar to \eqref{rf4a} and \eqref{rf4b} except that the interaction is now renormalized by the phase factors $e^{i{\bf Q}{\bf r}}$, i.e. $J({\bf r}) \to$ $J({\bf r})e^{i{\bf Q}{\bf r}}$.

In the equations \eqref{rf4a} and \eqref{rf4b} $B_{5/2}(5x/2)=[6\coth 3x - \coth (x/2)]/5$ is Brillouin function for spin 5/2 and $\phi^{-1}(M)$ is its inverse, $n=N/V$ is spins concentration i.e. number of spins per unit volume, $\beta=1/(k_BT)$ is reciprocal temperature. The equations \eqref{rf4a} and \eqref{rf4b} have a solution only for sufficiently large ratio $J_{zz}/J_{xy}$, where $J_{zz}$ and $J_{xy}$ are the amplitudes of AFM and FM exchange interactions \eqref{rf2}. As exchange interaction is short-range, to effectively create the long-range ordered AFM state, the concentration $n$ of Fe$^{3+}$ spins should be sufficiently large. This means that the concentration $n$ can also be regarded as parameter of transition between spin glass and magnetically ordered phase.   

If there is no dilution of Fe$^{3+}$ spin cites in above perovskites, the distribution function \eqref{rf3} becomes $\delta$ - function $f_{MF}({\bf H})=\delta({\bf H}-\beta H_0{\bf M})$ \cite{st97} and we have ordinary mean field approximation, where mean field $H_0=n\int J_{\alpha \beta}(\bf r)d^3r=$ $8\pi J_0nr_0^3$. It had been shown earlier \cite{ss02,ss03,st97} that regular procedure of transition from mean field to random field model is to expand the integrand of Eq. \eqref{rf4a} in power series in $\rho_{x,y,z}$. The second approximation, proportional to $\rho^2$, generates Gaussian distribution function of random fields. We note that our random field model in Gaussian approximation for Ising spins 1/2 gives ordinary replica-symmetric solution \cite{mpv} for spin glass. 

\begin{figure}
\begin{center}
\includegraphics [width=0.49\textwidth]{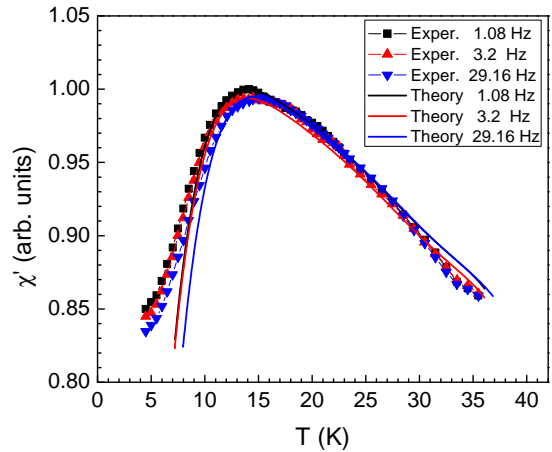}
\end{center}
\caption{(color online). Comparison of calculated (lines) and measured \cite{laguta13} (symbols)
temperature dependences (in the range of spin glass transition) of real part of magnetic susceptibility \eqref{tun1a}. Frequencies are shown in the legends. }\label{Fig2}
\end{figure}

\subsection{Dynamic magnetic susceptibility}

Our aim is to calculate experimentally observed $zz$ component of dynamic magnetic susceptibility 
$\chi_{zz}^*\equiv \chi^*$
\begin{equation}\label{tun1a}
\chi^*(H_{ext},T,\omega)=\frac{{\cal M}_z^*(H_{ext},T,\omega)}{H_{z, ext}},
\end{equation}
where ${\cal M}_z^*$ is $z$ component of dimensional magnetization (which at nonzero frequency $\omega$ becomes complex) and $H_{z, ext}$ is above external magnetic field. It can be shown (see \cite{my14} and references therein) that within above random field model the expression for dimensionless dynamic magnetization ${\bf M}^*(H_{ext},T,\omega)$ has the form
\begin{equation}\label{ds1}
M^*_{x,y,z}=\int \frac{H_{x,y,z}}{H}\ B_{5/2}\left(\frac 52 g\beta \mu_B H\right)\frac{f({\bf H},{\bf M})d^3H}{1+i\omega \tau({\bf H})}, 
\end{equation}
where $f({\bf H},{\bf M})$ is defined by Eq. \eqref{rf4a} and $\tau({\bf H})$ is a single spin relaxation time, averaged self-consistently over above random fields. Namely, assuming the Arrenius law for single spin relaxation, we obtain
\begin{equation}\label{ds2}
\tau={\bar \tau}\exp(-g\beta \mu_B {\bf S}{\bf H}),\ {\bar \tau}=\tau_0\exp(\beta U),
\end{equation}
where $U$ is a barrier between single spin orientation and $\tau_0\sim 10^{-12}$ sec is inverse attempt frequency or initial relaxation time. The parameters $U$ and $\tau_0$ are experimentally adjustable.
We note that while single spin relaxation in our model obeys Arrhenius law \eqref{ds2}, the relaxation of the whole spin ensemble at low temperatures $T \alt T_g$ obeys Vogel-Fulcher law \cite{ff1}, inherent for glassy systems. Latter law can be extracted from the resulting $\chi^*(H,T,\omega)$ curves only numerically.

\begin{figure}
\begin{center}
\includegraphics [width=0.47\textwidth]{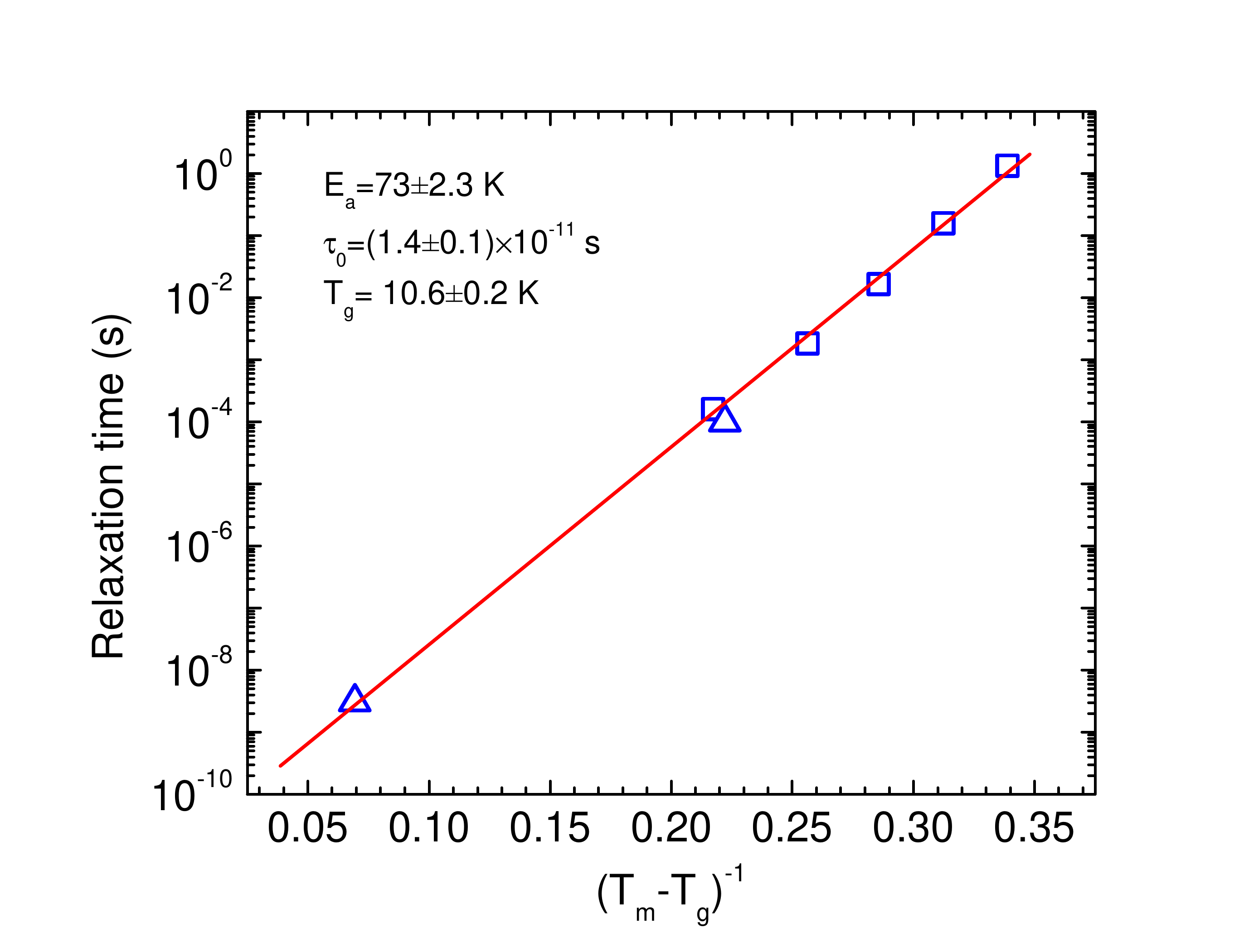}
\end{center}
\caption{(color online). Mean relaxation time as a function of reduced reciprocal temperature. Squares and triangles are experimental data derived from {\em {ac}} magnetic susceptibility and NMR relaxation respectively. The straight line is the fit to Eq. \eqref{ff1}. The parameters of fit are shown in the panel. }\label{Fig3}
\end{figure}

Explicit form of $\tau({\bf H})$ obtained with the help of Eq. \eqref{ds2} for spin 5/2 reads
\begin{equation}\label{ds3}
\tau({\bf H})={\bar \tau}\frac{\cosh 3X}{\cosh (X/2)},\ X=g\beta\mu_B(H+H_{ext}).
\end{equation}

The expression \eqref{tun1a} with respect to \eqref{ds1}, \eqref{rf4a} and \eqref{ds3} permits to determine theoretically the dynamic magnetic susceptibility of considered disordered perovskite multiferroics. 

\section{Comparison with experiment}
\begin{figure}
\begin{center}
\includegraphics [width=0.5\textwidth]{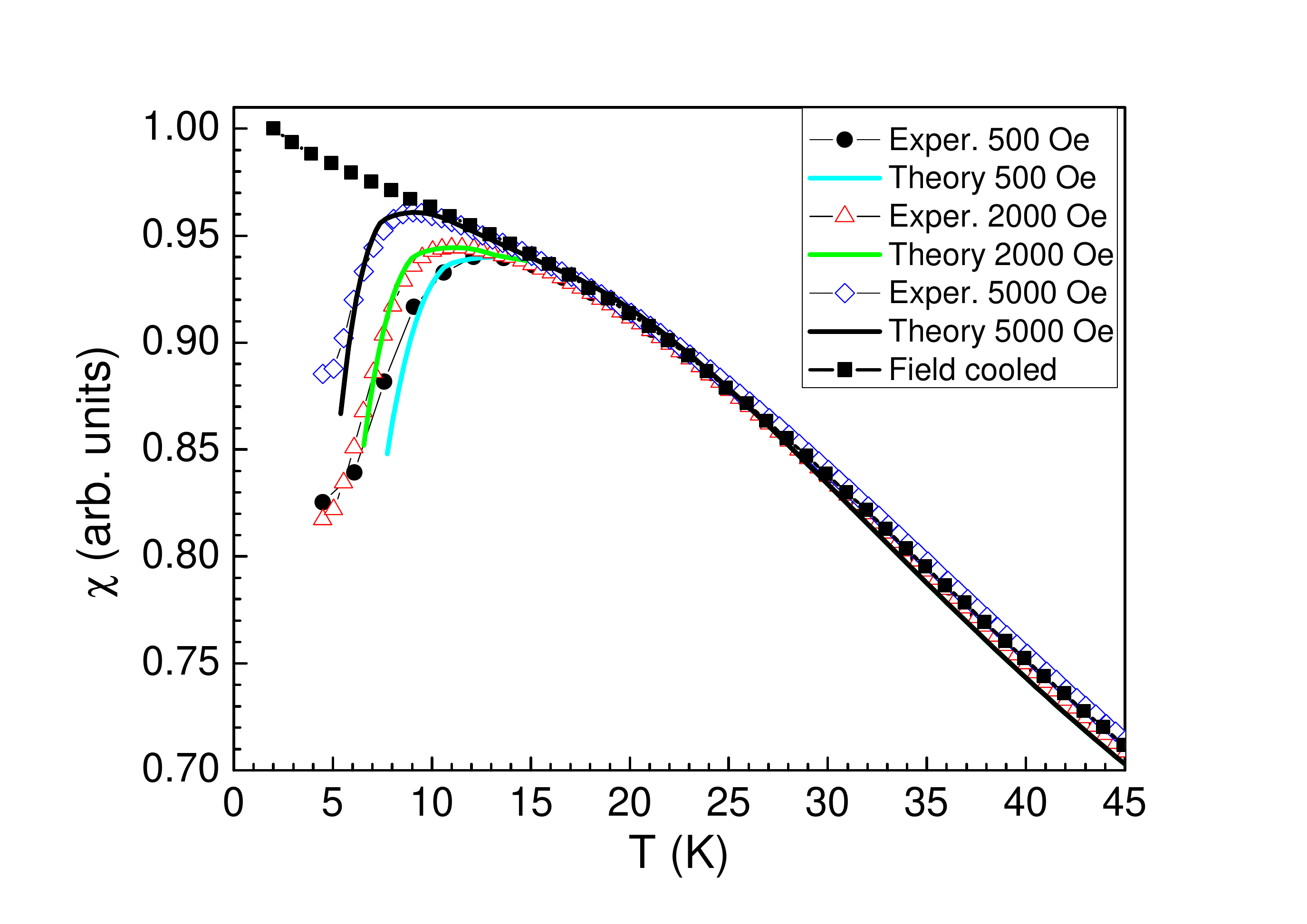}
\end{center}
\caption{(color online). Comparison of calculated (lines) and measured \cite{laguta13} (symbols)
temperature dependences of ZFC magnetic susceptibility at different magnetic fields shown in the legend. Experimental FC susceptibility is shown for comparison.}\label{Fig4}
\end{figure}

Figure \ref{Fig2} reports the fit of the {\em {ac}} $\chi'(\tau)$ to experimental curves measured in \cite{laguta13} for PFN at different frequencies. The excellent coincidence between theory and experiment is seen. We fit the experimental curves by their maximum temperatures $T_m$, which varies from 13.8 to 14.4 K with frequency increase from 1.08 to 29.16 Hz. Corresponding mean relaxation times of spin fluctuations as a function of the $T_m$ temperature are reported in Fig. \ref{Fig3}. To extract the experimental symbols shown in Fig. \ref{Fig3}, we use both relaxation time data from the {\em{ac}} magnetic susceptibility and those from nuclear magnetic resonance (NMR) of $^{17}$O isotope. It has been shown previously \cite{blinc08} that both spin-lattice ($T_1$) and spin-spin ($T_2$) nuclear magnetization relaxation times show distinct minimum around the SG transition temperature. In particular, the $T_1$ relaxation time depends on electron spin fluctuations at the nuclear Larmor frequency $\omega/(2\pi)=51.52$ MHz, while the $T_2$ relaxation time feels spin fluctuations at much lower frequency, defined by the spin-echo delay time $2\tau_s=$ 100 $\mu$s. The NMR data thus extend the measured relaxation times of the spin fluctuations up to nanosecond range. The data in Fig. \ref{Fig3} are described well by the Vogel - Fulcher law \cite{ff1} at the time scale from 1 down to $3\cdot 10^{-9}$ s:
\begin{equation}\label{ff1}
\frac{1}{2\pi f}=\tau_0\exp\left[\frac{E_a}{T_m-T_g}\right],
\end{equation} 
where $T_m$ corresponds to the peak temperature either of {\em{ac}} magnetic susceptibility or nuclear magnetic relaxation at frequency $f$.

Fig. \ref{Fig4} shows the fit of the dc magnetic susceptibility (measured at field cooling (FC) and zero-field cooling (ZFC) protocols) as a function of temperature for different strength of the field.  The excellent coincidence between theory and experiment is seen again. Note that with increase of magnetic field strength the $T_m$ temperature at the ZFC curves shifts to lower temperature in accordance with our theory prediction, see also Ref. \onlinecite{gt81}. The above shift is related to 
the interplay between spins polarization by the applied magnetic field and transversal spin components freezing, giving rise to the SG phase. Namely, at higher magnetic fields the polarized (i.e. aligned along field direction) spins form effectively paramagnetic phase so that lower temperatures are required for transversal spin components to "feel" the glassy correlations i.e. to freeze into SG phase. 

Our fitting of experiment in Figs. \ref{Fig2} and \ref{Fig4} has been done for the following set of parameters. Considering the magnetic moment of Fe$^{3+}$ ion to be 5.2 $\mu_B$ ($\mu_B$ is Bohr magneton) and taking the best fit value $nr_0^3=0.3$, we obtain that concentration of Fe$^{3+}$ ions is approximately 7.8$\cdot 10^{21}$ cm$^{-3}$. The barrier $U$ between single spin orientation in Arrhenius law \eqref{ds2} for our fit turns out to be $U=350$ K. Likewise, the parameter $\tau_0=1.47 \cdot 10^{-12}$ sec. We point here to the difference between parameters of single-spin relaxation Arrhenius law and those for the Vogel - Fulcher law \eqref{ff1}, which results from the average relaxation of the whole spins ensemble. Also, the best fit is achieved for the exchange constants ratio $|J_{xy}/J_{zz}|\approx 0.05$, see Eq. \eqref{rf2}.

The discrepancy between theoretical and experimental curves at low temperatures are due to the fact that additional defects contribute experimental susceptibility, making it to decay slower. The pretty good coincidence between theoretical and experimental magnetic susceptibility curves at different frequences and magnetic fields suggests that the observed behavior can be well attributed to the joint action of site disorder, anisotropy (when $xy$ components of spin vector fluctuate making the length of $z$ component to vary) and hierarchy of exchange interactions in PFN. Namely, although the short-range exchange interaction between spins dominates, the smaller frustrating exchange interaction in the $xy$ plane forms the observable SG anomalies owing to transversal spin fluctuations. 

\section{Conclusions}
To conclude, here we present the experimental explanation and its quantitative theoretical description 
of the coexistence of long-range ordered (AFM) and spin-glassy phases in the crystalline double perovskite Fe - contained multiferroics like PbFe$_{1/2}$Nb$_{1/2}$O$_3$. The main physical mechanism of such coexistence is the interplay between chemical disorder (half of corresponding lattice cites are occupied by spinless Nb$^{5+}$ ions) and the anisotropy and frustration of exchange interactions between Fe$^{3+}$ spins in above multiferroics and PFN in particular. Namely, if the AFM long-range order is related to the strong exchange interaction of $z$ - components of spins, the glassy effects are due to the much weaker exchange interaction of ferromagnetic sign, which couples transversal $xy$ spin components. Our theory, explicitly considering randomness in the spin space along with anisotropy of exchange interactions, is able to describe our experimental data quantitatively thus showing that considered physical mechanism of AFM and glassy phases coexistence in above Fe - contained multiferroics is quite accurate.    

\acknowledgements
VVL acknowledges financial support from the GACR (project 13-11473S).

\end{document}